\documentstyle[aps,prd,eqsecnum]{revtex}
\begin{document}
\draft
\title{Observable Algebra}
\author{ Merab Gogberashvili}
\address{ Andronikashvili Institute of Physics \\
6 Tamarashvili Str.,Tbilisi 380077, Georgia \\
{\sl E-mail: gogber@hotmail.com } }
\twocolumn[\hsize\textwidth\columnwidth\hsize\csname
           @twocolumnfalse\endcsname
\date{\today}
\maketitle
\widetext
\begin{abstract}
A physical applicability of normed split-algebras, such as
hyperbolic numbers, split-quaternions and split-octonions is
considered. We argue that the observable geometry can be described
by the algebra of split-octonions. In such a picture physical
phenomena are described by the ordinary elements of chosen
algebra, while zero divisors (the elements of split-algebras
corresponding to zero norms) give raise the coordinatization of
space-time.  It turns to be possible that two fundamental
constants (velocity of light and Planck constant) and uncertainty
principle have geometrical meaning and appears from the condition
of positive definiteness of norms. The property of
non-associativity of octonions could correspond to the appearance
of fundamental probabilities in four dimensions. Grassmann
elements and a non-commutativity of space coordinates, which are
widely used in various physical theories, appear naturally in our
approach.
\end{abstract}
\medskip
\noindent{}
\medskip
]
\narrowtext

\section{Introduction}

Real physical phenomena always expose themselves experimentally as
a set of measured numbers. In a general way one can say that the
measurement process (even simple counting) consists of matching
the physical phenomenon to be measured with some familiar set of
objects, or the number words. The theory of representations of
algebras serves as a tool that gives the possibility of
interpreting abstract mathematical quantities from the point of
view of the observable reality. The geometry of space-time, being
the main characteristic of the reality, can be described also in
the language of algebras and symmetries.

Usually in physics the geometry is thought to be objective without
any connection to the way it is observed. One assumes the
existence of some massive objects, with respect to which physical
events can be numerated. It is known that physical limits on
masses of reference systems cause large uncertainties in the
measurement even of a large distances \cite{NgDa}. In general
geometry cannot be introduced independently. Results of any
observation depend not only upon the phenomena we study but also
on the way we make the observation and on the algebra one uses for
a coordinatization. The properties of space-time (such as
dimension, distance, etc.) do not exist separately, but are just
reflections of universal symmetries of the algebra using in the
measurement process.

Let us start from the assumption that physical quantities must
correspond to the elements of some algebra, so they can be summed
and multiplied. Since all observable quantities we extract from
single measurements are real, it is possible to restrict ourselves
to the field of real numbers. To have a transition from a manifold
consisting the results of measurements to geometry (by
multiplication of elements corresponding to the direct and reverse
signals), one must be able to introduce a distance between two
objects. Introduction of a distance always means some comparison
of two objects using one of them as an etalon. Thus we need an
algebra with a unit element.

In the algebraic language, these requirements mean that to
describe the geometry of the real world we need a normed
composition algebra with the unit element over the real numbers.
Besides of the usual algebra of real numbers there are, according
to the Hurwitz theorem, three extraordinary and unique algebras
that satisfy the required conditions. These are the algebras of
complex numbers, quaternions and octonions \cite{Sc}.

Essential for all these algebras is the existence of the real unit
element $e_0$ and a different number of adjoined hyper-complex
units $e_n$. For the case of complex numbers $n = 1$, for
quaternions $n = 3$ and for octonions $n = 7$. The square of the
unit element $e_0$ is always positive and the squares of the
hyper-complex units $e_n$ can be positive or negative
\begin{equation} \label{e-pm}
e_n^2 = \pm e_0 ~.
\end{equation}
In applications of division algebras mainly the negative sign in
(\ref{e-pm}) is used. In this case norm of the algebra is
positively defined.

We note that in real world any comparison of two objects can be
done only by the exchange of some signals between them. One-way
signals can correspond to any kind of number. The important
requirement is to have a real norm. We assume that the norm, the
result of one measurement, is actually the multiplication of
direct and backward signals. So the new physical requirement on
the algebra we would like to apply is that it must contain special
class of elements corresponding to the unit signals (which can be
used for measurements of distances). For this purpose we want to
choose the positive sign in (\ref{e-pm}), or
\begin{equation} \label{e-}
e_n^2 = e_0 ~.
\end{equation}
This leads to so-called split algebras with the equal number of
terms with the positive and negative signs in the definition of
their norms. Adjoining to a real unit element (which could be
connected with time), any invertible hyper-complex units with
negative norms could be understood as an appearance of dynamics.
The critical elements of split-algebras, so-called zero divisors,
corresponding to zero norms can be used as the unit signals to
measure distances between physical objects, wile ordinary physical
objects are assumed to be described by regular elements (with
division properties) of the algebra.

We believe that the dimension of the proper algebra and the
properties of its basis elements also can be extracted from
physical considerations.

It is known that observed physical quantities are divided into
three main classes, scalars, vectors, and spinors. We need to have
a unique way to describe these objects in an appropriate algebra.

Scalar quantities have magnitude only, and do not involve
directions. The complete specification of a scalar quantity
requires a unit of the same kind and a number counting how many
times the unit is contained in the quantity. Scalar quantities are
manipulated by applying the rules of ordinary algebra or real
numbers.

A vast range of physical phenomena finds its most natural
description as vector quantities, being a way to describe the size
and orientation of something. A vector quantity requires for its
specification a real number, which gives the magnitude to the
quantity in terms of the unit (as for scalars) and an assignment
of direction. By revising the process of addition of physical
vectors it was found that using unit orthogonal basis vectors
$e_n$ any vector could be decomposed into components
\begin{equation} \label{a}
{\bf a} = a_ne_n ~.
\end{equation}
The ordinary idea of a product in scalar algebra cannot apply to
vectors because of their directional properties. Since vectors
have their origin in physical problems, the definition of the
products of vectors historically was obtained from the way in
which such products occur in physical applications. Two types of
multiplication of vectors were introduced - the scalar product
\begin{equation} \label{scalar}
({\bf ab}) = ab\cos{\theta} ~,
\end{equation}
and the vector product
\begin{equation} \label{vector}
 [{\bf ab}] = {\bf n}~ab\sin{\theta} ~,
\end{equation}
where ${\bf n}$ is a unit vector perpendicular to the plane of
${\bf a}$ and ${\bf b}$.

Formulae (\ref{a}) - (\ref{vector}) result in the following
properties of unit basis vectors
\begin{eqnarray} \label{e-vector}
(e_ne_m) = 0 ~, ~~~~~ e_ne_m = - e_me_n ~, \nonumber\\
(e_ne_n) = e_n^2 = 1~, ~~~[e_ne_m] = \varepsilon_{nmk}e_k ~,
\end{eqnarray}
where $\varepsilon_{nmk}$ is fully anti-symmetric tensor. Or vice
versa, we can say that from the postulated algebra
(\ref{e-vector}) for basis elements one recovers the observed
multiplication laws (\ref{scalar}) and (\ref{vector}) of physical
vectors.

Because of the ambiguity in choosing of direction of basis
vectors, one must introduce conjugate elements $e_n^*$, which
differ from $e_n$ by the sign
\begin{equation} \label{e*}
e_n^* = - e_n~,
\end{equation}
what physically can be understood as a reflection.

The algebra of the vector basis elements (\ref{e-vector}) and
(\ref{e*}) is exactly the same as we have for basis units of split
algebras (with the positive sign (\ref{e-})), which we had chosen
from some general requirements. Since only vector products of
different basis vectors and also scalar product of a basis vector
with itself are non-zero, we do not need to write different kind
of brackets to express their multiplications.

It is known that the notions of vector products and composition
algebras in fact are equivalent \cite{Ro}. However, pure
quaternions (with the negative sign in (\ref{e-pm})) can not
produce the usual 3-dimensional vectors, since with the respect of
reflections quaternions basis units behave like a pseudo-vectors
\cite{SiMa}. Generalization of the standard vector products is
only possible in seven dimensions and is generated by octonions
\cite{Ro,Si}. We shall show below that exactly three orthogonal
vector-like objects are elements of the split-octonion algebra.

Restrictions on the appropriate algebra can be imposed also from
the properties of spinors, another important class of observed
physical quantities. Spinors as vectors can be characterized by
some magnitude and direction, but they changes sign after a $2\pi
$ rotation in 3-space. Using a multi-dimensional picture, when we
transform ordinary space coordinates, spinors rotate also in some
extra space. It is well known that there exists an equivalence of
vectors and spinors in eight dimensions \cite{CaGa}, which renders
the 8-dimensional theory (connected with octonions) a very special
case. In eight dimensions a vector and a left and right spinor all
have an equal number of components (namely, eight) and their
invariant forms all look the same. One may consider the vector's
rotation as the primary rotation that induces the transformations
of two kinds of spinors, or equivalently starts from the spinor
rotation, which induces a corresponding rotation of the vector and
the second kind of spinor. This property is referred in the
literature as "principle of triality". The physical result of this
principle can be the observed fact that interaction in particle
physics exhibit by the vertex containing a left spinor, a right
spinor and a vector \cite{Ba}.

To summarize our general considerations, we conclude that the
proper algebra needed to describe the real world must be the
8-dimensional algebra of split-octonions over the real numbers.
Geometry can be constructed by the exchanging of octonionic
signals.

The assumption that our Universe is built from pairs of octonions
is now an important idea of string theory also (for example, see
some recent publications \cite{string}). String theory (where
octonions are used only for extra dimensions), however, in
difference with our approach is an attempt to marry classical
considerations about 4-dimensional space-time to the small-scale
uncertainties inherent in quantum theory.

In the next section the properties of zero divisors, which appear
in split-algebras, are considered. The following three sections
are devoted to geometrical applications of hyper-numbers,
split-quaternions and split-octonions, respectively. Since
physical applications of split-algebras over the real numbers are
purely studied main considerations of all sections are original.


\section{Zero Divisors}

The division and split algebras have some differences following
from the fact that in split-algebras special objects, called zero
divisors, can be constructed \cite{Sc,SoLo}. We assume that these
objects (we shall call them critical elements of the algebra)
could be serve as the unit signals characterizing physical events.
Sommerfeld specially notes physical importance of zero divisors of
split algebras in his book \cite{So}.

The first types of zero divisors we want to study are projection
operators with the property
\begin{equation} \label{D}
D^2 = D ~,
\end{equation}
for any non-zero $D$. In division algebras only projection
operator is the identity $1 = e_o$.

The most general form of projection operator in split-algebra is
\begin{equation} \label{anen}
D = \frac{1}{2} + a_n e_n ~,
\end{equation}
with $a_na_n = 1/4$. In (\ref{anen}) index $n$ runs over the
number of the orthogonal hyper-complex basis elements $e_n$.

In general, a zero divisor is called primitive if it cannot be
expressed as the sum of two other commuting zero divisors. For
simplicity below we shall use only primitive projection operators.

Commuting primitive projection operators can be constructed by
decomposition of the identity element
\begin{equation} \label{1}
1 = D^+_n + D^-_n ~.
\end{equation}
with the particular choice
\begin{equation} \label{D=pm}
D^+_n = \frac{1}{2}(1 + e_n)~, ~~~D^-_n = \frac{1}{2}(1 - e_n) ~.
\end{equation}
The elements $D^+_n$ and $D^-_n$ commute, since
\begin{equation} \label{commuting}
D^+_nD^-_n = D^-_nD^+_n = 0 ~.
\end{equation}

Another type of zero divisors in split-algebras are Grassmann
numbers defined as the set of anti-commuting numbers ${G_1, G_2,
...,G_n }$ with the properties
\begin{equation} \label{G}
G^2_n = 0~, ~~~~~{G_n G_m} = 0~.
\end{equation}
Since the square of hyper-complex basis units of a split-algebra
is $+1$ or $-1$, Grassmann numbers can be constructed by coupling
of two basis elements (except of unity) with the opposite choice
of the sign in (\ref{e-pm}). For example, primitive Grassmann
numbers are the sums of any two basis elements $e_1$ and $e_2$
\begin{equation} \label{G-pm}
G^\pm = \frac{1}{2}(e_1 \pm e_2)
\end{equation}
with the properties
\begin{equation} \label{e12}
e_1^2 = 1 ~, ~~ e_2^2 = -1 ~, ~~ e_1^* = - e_1 ~, ~~ e_2^* = -
e_2~.
\end{equation}

A Grassmann number does not contain the unit element $1$ as the
one of its terms (the structure with the unit element forms
projection operator) and can be represented as
\begin{equation} \label{G=D}
G = e_1 D ~,
\end{equation}
where $D$ is a projection operator and $e_1$ is a vector-like
basis element of the algebra with positive square $e_1^2 = 1$.

For the completeness we note that the Grassmann numbers $G^\pm $
and projection operators $D^\pm$ obey the following algebra:
\begin{eqnarray} \label{GD}
D^\pm D^\mp = 0 ~, ~~~~~ D^\pm D^\pm = D^\pm ~, \nonumber \\
G^\pm G^\pm = 0 ~, ~~~~~ G^\pm G^\mp = D^\mp ~, \nonumber \\
D^\pm G^\pm = 0 ~, ~~~~~ D^\pm G^\mp = G^\mp ~, \\
G^\pm D^\mp = 0 ~, ~~~~~ G^\pm D^\pm = G^\pm ~. \nonumber
\end{eqnarray}
From this relations we see that, separately, the quantities $G^+$
and $G^-$ are Grassmann numbers, but they do not commute with each
other (in contrast with the projection operators $D^+$ and $D^-$)
and thus do not obey the full Grassmann algebra (\ref{G}). Instead
the quantities $G^+$ and $G^-$ are the elements of so-called
algebra of Fermi operators with the anti-commutator
\begin{equation} \label{fermi}
G^+G^- + G^-G^+ = 1~.
\end{equation}
The algebra of Fermi operators is some syntheses of the Grassmann
and Clifford algebras.

Commuting zero divisors correspond to simultaneously measurable
signals and can serve as the geometrical basis for the physical
events. Ordinary physical events are describing by the regular
elements of the algebra.


\section{Hyper-Numbers}

There are certain important types of physical phenomenon that are
intrinsically 2-dimensional in nature. Many physical functions
occur naturally in pairs and are expressed by 2-dimensional real
numbers that satisfies a different type of algebra \cite{StNe}.
The most common 2-dimensional algebra is the algebra of complex
numbers, however, there is no special reason to prefer one algebra
to the other \cite{Sa}.

Historically, the motivation to introduce complex numbers was
mathematical rather than physical. This numbers made sense of the
solution of algebraic equations, of the convergence of series, of
formulae for trigonometric functions, differential equations, and
many other things. This was quite unlike the initial motivation
for using real numbers, which came about as an idealization of the
kind of quantity that directly arose from physical measurements.

Essential for the complex numbers
\begin{equation} \label{z}
z= x + iy ~,
\end{equation}
where $x$ and $y$ are some real numbers, is existence of the unit
element and one imaginary element $i$, with the property $i^2 =
-1$. Later it was found that if we introduce the conjugation of
complex numbers
\begin{equation} \label{z*}
z^* = x - iy ~,
\end{equation}
there algebra is division. A regular complex number can be
represented geometrically by the amplitude and polar angle
\begin{equation} \label{Nz}
\rho^2 = zz^* = x^2 + y^2 ~, ~~~~ \theta = \arctan{\frac{y}{x}} ~.
\end{equation}
The amplitude $\rho $ is multiplicative and the polar angle
$\theta$ is additive upon the multiplication of complex numbers.

Complex number have found many uses in physics, but these were
considered as a "mathematical tricks", like the employment of
complex numbers in 2-dimensional hydrodynamics, electrical circuit
theory, or the theory of vibrations.

In quantum mechanics, complex numbers enter at the foundation of
the theory from probability amplitudes and the superposition
principle. To pass from a quantum information link, which does not
respect the rules of ordinary classical causality (as in the case
of Einstein-Podolsky-Rosen phenomenon) to a classical information
link (which necessarily propagates causally into the future) and
to obtain real number corresponding to the probability, one must
multiply the complex amplitude by its complex conjugate. In fact,
the complex conjugated amplitude may be thought of as applying to
the quantum-information link in the reverse direction in time
\cite{Pe}.

Up to now it is admitted without justification that two-valuedness
of physical quantities naturally appearing in quantum mechanics is
to be described in terms of complex numbers. Complex numbers
achieve a particular representation of quantum mechanics in term
of which the fundamental equations take their simplest form. Other
choices for the representation of the two-valuedness also
possible, but would give to the Schrodinger equation a more
complicated form (involving additional terms), although its
physical meaning would be unchanged \cite{CeNo}.

The main reason why complex numbers are popular is Euler's
formula. As it was mentioned by Feynman \cite{Fe}, "the most
remarkable formula in mathematics is:
\begin{equation}
e^{i\theta} = \cos{\theta} + i\sin{\theta} ~. \nonumber
\end{equation}
This is our jewel. We may relate the geometry to the algebra by
representing complex numbers in a plane
\begin{equation} \label{euler}
x + iy = \rho e^{i\theta} ~.
\end{equation}
This is the unification of algebra and geometry."

Using (\ref{euler}) the rule for complex multiplication looks
almost obvious as a consequence of the behavior of rotations in
plane. We know that a rotation of $\alpha$-angle around the
$z$-axis, can be represented by
\begin{equation}
e^{i\alpha}(x + iy) = \rho e^{i(\theta + \alpha )} ~. \nonumber
\end{equation}

Another possible 2-dimensional normed algebra is the algebra of
hyper-numbers
\begin{equation} \label{zh}
z = T + hx~.
\end{equation}
This has also a long history but is rarely used in physics
\cite{OlAnFj}. In (\ref{zh}) the quantities $T, x$ are the real
numbers and hyper-unit $h$ has the properties similar to ordinary
unit vector
\begin{equation} \label{h2}
h^2 = 1~.
\end{equation}

Conjugation of $h$ can be understand as reflection when $h$
changes its direction on the opposite $h^* = -h$. So, in contrast
with ordinary complex numbers, the norm of a hyper-number is
invariant under of rotation with reflection.

Hyper-numbers constitute a commutative ring, but not a field,
 since norm of a nonzero hyper-number
\begin{equation} \label{Nzh}
N = zz^* = T^2 - x^2 ~,
\end{equation}
can be zero.

A hyper-number does not have an inverse when its norm is zero, i.
e., when $x = \pm T$, or, alternatively, when $z = T(1 \pm h)$.
These two lines whose elements have no inverses, play the same
role as the point $z = 0$ does for the complexes, and provide the
essential property of the light cone, which makes the
hyper-numbers relevant for the representation of relativistic
coordinate transformations. Lorentz boosts can be succinctly
expressed as
\begin{equation} \label{lorentz}
(T + hx) = (T' + hx')e^{h\theta}~,
\end{equation}
where $\tanh{\theta} = v$.

Any hyper-number (\ref{zh}) can be expressed in terms of light
cone coordinates
\begin{equation} \label{zhDec}
z = T +hx = (T + x)D^+ + (T - x)D^- ~,
\end{equation}
where we introduced projection operators
\begin{equation} \label{Dzh}
D^\pm = \frac{1}{2}(1 \pm h)
\end{equation}
corresponding to the critical signals.

The unit of the algebra of hyper-numbers is connected with the
critical signals and is universal for all physical quantities
(which are described by the ordinary elements of the algebra).
Thus it is just the hyper-unit is responsible for the signature of
the space-time metric and the velocity of light $c$ is related
with it. Thus the norm (\ref{Nzh}) can be understand as the
distance and the constant $c$ explicitly extracted from the unit
element of the algebra
\begin{equation} \label{Nzch}
N = c^2t^2 - x^2 ~,
\end{equation}
where quantity $t$ now has the dimension of time. So we can say
that hyper-numbers are useful to describe dynamics in (1+1)-space.


\section{Quaternions}

William Hamilton's discovery of quaternions in 1843 was the first
time in history when the concept of 2-dimensional numbers was
successfully generalized. Many authors have proposed applications
of quaternions in physics \cite{Sa,Qua}.

General element of the quaternion algebra can be written by using
only the two basis elements $i$ and $j$ in the form
\begin{equation} \label{q}
q = a + bi + (c + di)j ~,
\end{equation}
where $a, b, c$ and $d$ are some real numbers. The third basis
element of the quaternion $(ij)$ is possible to obtained by
composition of the first two.

The quaternion reverse to (\ref{q}), the conjugated quaternion
$q^*$, can be constructed using the properties of the basis units
under the conjugation (reflection)
\begin{equation} \label{*}
i^* = - i~, ~~~j^* = - j~, ~~~ (ij)^* = - (ij) ~.
\end{equation}
A physically observable quantity is the norm of a quaternion $N =
qq^*$ corresponding to the multiplication of the direct and
reverse signals.

When the basis elements $i$ and $j$ are imaginary $ i^2 = j^2 = -1
$ (similar to ordinary complex unit) we have Hamilton's quaternion
with the positively defined norm
\begin{equation} \label{Nq+}
N = qq^* = q^*q = a^2 + b^2 + c^2 +d^2 ~.
\end{equation}
In this case the third orthogonal basis unit $(ij)$ has properties
analogous to $i$ and $j$ (they all are similar to pseudo-vectors
\cite{SiMa}). Because of this fact it is impossible to identify
pure quaternion with a vector \cite{SiMa,Qua}.

For the positive squares $i^2= j^2 = 1$ we have the algebra of
split-quaternions. The unit elements $i, j$ have the properties of
real unit vectors. The norm of a split-quaternion
\begin{equation} \label{Nq-}
N = qq^* = q^*q = a^2 - b^2 - c^2 + d^2 ~,
\end{equation}
has (2+2)-signature and in general is not positively defined.

We see that the third unit element $(ij)$ of split-quaternions is
the vector product of the unit vectors $i$ and $j$ and thus is a
pseudo-vector. It is important to realize that a pseudo-vector is
not a geometrical object in the usual sense. In particular, it
should not be considered as a real physical arrow in space.

The pseudo-vector $(ij)$ differs from the other two hyper-complex
units of split-quaternions $i$ and $j$ and behaves like a pure
imaginary object
\begin{equation} \label{ij}
(ij)^* = - (ij) ~, ~~~ (ij)(ij) = - i^2j^2 = -1 ~.
\end{equation}
In this fashion one can justify the origins of complex numbers
\begin{equation} \label{z1}
z = x + (ij)y~,
\end{equation}
without introducing them ad-hoc. In this sense we can say that
split-quaternions are more rich in structure compared to
Hamilton's quaternions, since they contain complex and
hyper-complex numbers as particular cases.

In the algebra of split quaternions two classes (totally four)
projection operators
\begin{equation} \label{Dq}
D^\pm_i = \frac{1}{2}(1 \pm i) ~, ~~~ D^\pm_j = \frac{1}{2}(1 \pm
j)~,
\end{equation}
can be introduced. The two classes does not commute with each
other. The commuting projection operators with the properties
\begin{equation} \label{D12}
[D^+D^-] = 0 ~, ~~ D^+D^+ = D^+~, ~~ D^-D^- = D^-~,
\end{equation}
are only the pairs $D^\pm_i$, or $ D^\pm_j$. The operators $D^+$
and $D^-$ differ from each other by reflection of the one basis
element and thus correspond to the direct and reverse critical
signals along one of the two real directions.

In the algebra we have also two classes of Grassmann numbers
\begin{equation} \label{Gq}
G^\pm_i = \frac{1}{2}(1 \pm i)j ~, ~~~ G^\pm_j = \frac{1}{2}(1 \pm
j)i ~.
\end{equation}

From the properties of zero divisors
\begin{eqnarray} \label{DGq}
D^\pm_i G^\pm_i = G^\pm_i ~, ~~~ D^\pm_j G^\pm_j = G^\pm_j ~, \nonumber\\
D^\pm_i G^\mp_i = 0 ~, ~~~~~~ D^\pm_j G^\mp_j = 0 ~,
\end{eqnarray}
we see that the Grassmann numbers (\ref{Gq}) have pairwise
commuting relations with the projection operators (\ref{Dq})
\begin{equation} \label{dg}
[D^\pm_i G^\mp_i] = 0 ~, ~~~ [D^\pm_j G^\mp_j] = 0 ~.
\end{equation}

Using commuting zero divisors any quaternion can be written in the
form
\begin{eqnarray} \label{qDec}
q = a + i b +(c + id)j = \nonumber\\
 = D^+ [a + b + (c + d)G^+] + \\
 + D^- [a - b +(c - d)G^-] ~,\nonumber
\end{eqnarray}
where $D^\pm$ and $G^\pm$ are the projection operators and
Grassmann elements belonged to the one of the classes from
(\ref{Dq}) and (\ref{Gq}) labelled by $i$, or by $j$.

The quaternion algebra is associative and therefore can be
represented by matrices. We get the simplest non-trivial
representation of the split-quaternion basis units if we choose
the real Pauli matrices accompanied by the unit matrix. Note that
for a real matrix representation of Hamilton's quaternions one
needs 4-dimensional matrices.

The independent unit basis elements of split quaternions $i$ and
$j$ have the following matrix representation
\begin{equation}
i =\pmatrix{1 & 0 \cr 0 & -1 }~, ~~~ j = \pmatrix{0 & 1 \cr 1 &
0}~.
\end{equation}
The third unit basis element $(ij)$ is formed by multiplication of
$i$ and $j$ and has the representation
\begin{equation}
(ij) =\pmatrix{0 & 1 \cr -1 & 0 }~.
\end{equation}
It is easy to noticed that, in contrast with the complex case, the
three real Pauli matrices have different properties, since their
squares give the unit matrix
\begin{equation}
(1) =\pmatrix{1 & 0 \cr 0 & 1 }
\end{equation}
with different signs
\begin{equation} \label{square}
i^2 = j^2 = (1) ~, ~~~~~(ij)^2 = -(1) ~.
\end{equation}

Conjugation of the unit elements $i$ and $j$ means changing of
signs of matrices $i$, $j$ and $(ij)$, thus
\begin{equation} \label{i-norm}
ii^* = jj^* = -(1) ~, ~~~~~(ij)(ij)^* = (1) ~.
\end{equation}

Matrix representation of the independent projection operators and
Grassmann elements from (\ref{Dq}) and (\ref{Gq}) labelled by $i$
are
\begin{eqnarray}
D^+_i =\frac{1}{2}(1+i) =\pmatrix{1 & 0\cr 0 & 0 } ~,\nonumber
\\
D^-_i =\frac{1}{2}(1-i) =\pmatrix{0 & 0\cr 0 & 1 } ~,\nonumber
\\
G^+_i =\frac{1}{2}(j+ij) = \pmatrix{0 & 1\cr 0 & 0 } ~,\\
G^-_i =\frac{1}{2}(j-ij) = \pmatrix{0 & 0\cr 1 & 0 } ~.\nonumber
\end{eqnarray}
It is easy to find also matrix representation of similar zero
divisors labelled by the second index $j$.

The decomposition (\ref{qDec}) of a quaternion  now can be written
in the form
\begin{equation}
q = \pmatrix{(a+b) & (c+d) \cr (c-d) & (a-b) }
\end{equation}
and the norm (\ref{Nq-}) is the determinant of this matrix
\begin{equation} \label{q-norm}
{\it det} q = (a^2 - b^2) - (c^2 - d^2) ~.
\end{equation}

If we treat the unit element of the algebra as the time, then
split-quaternions could be used for describing of dynamics in
2-dimensional space, just as hyper-numbers are used to study
dynamics in 1-space. Here we have two independent rotations in the
$(t-i)$ and $(t-j)$ planes described by $D^\pm$, which similar to
hyper-numbers case (\ref{Nzch}) introduces fundamental constant
$c$.

But what is the geometrical meaning of the Grassmann numbers and
the fourth orthogonal element in the split-quaternion algebra?
There was no similar element in the algebra of hyper-numbers.

It can be shown, that pseudo-vector $(ij)$ is connected with the
rotation in the $(i-j)$ plane. Indeed, the operators
\begin{equation} \label{Rq}
R^\pm = \frac{1}{\sqrt{2}}[1 \pm (ij)] ~,
\end{equation}
(which are ordinary complex numbers with the property $ R^+R^- =
1$) represent the rotation by $\pi /2$ in $(i-j)$ plane. Thus they
can be used to transform $i$ into $j$
\begin{equation} \label{i-j}
R^+iR^- = - j ~, ~~~~~ R^+jR^- = i~.
\end{equation}
Thus the extra pseudo-direction $(ij)$ is similar to the angular
momentum and it is natural to assume that it can be connected with
the property of the matter such as spin.

In the algebra there is a symmetry in choice of the independent
projection operators $D_i^\pm$ or $D_j^\pm$ for the quaternion
decomposition (\ref{qDec}). This means that we have two
independent space directions $i$ and $j$, where in principle
signals could be exchanged. However, we can simultaneously measure
signals only from the one direction (since the operators $D_i^\pm$
and $D_j^\pm$ do not commute). This property is equivalent to the
non-commutativity of coordinates \cite{Non}.

The constant $c$, as for hyper-numbers (\ref{Nzch}), can be
extracted explicitly from the unit element of any quaternion
\begin{equation} \label{dual}
q = ct + xi + yj + \lambda (ij) ~,
\end{equation}
where $\lambda$ is some quantity with the dimension of length.

Projection operators correspond to critical rotations for the
$(1-i)$ and $(1-j)$ planes and introduce maximal velocity,
following from the requirement to have the positive norm for the
quaternion (\ref{dual})
\begin{equation} \label{delta-x}
\frac{\Delta x}{c\Delta t} < 1 ~, ~~~~~ \frac{\Delta y}{c\Delta t}
< 1 ~.
\end{equation}
Analogously Grassmann numbers correspond to critical rotations in
the $(i-ij)$ and $(j-ij)$ planes. When conditions (\ref{delta-x})
are satisfied for the positive definiteness of the norm, now we
must write also the extra relations
\begin{equation} \label{delta-l}
\frac{\Delta x}{\Delta \lambda} > 1 ~, ~~~~~ \frac{\Delta
y}{\Delta \lambda} > 1 ~.
\end{equation}
This means that there must exist the second fundamental constant
(which can be extracted from $\lambda$) characterizing this
critical property of the algebra. Since the pseudo-direction
$(ij)$ is similar to angular momentum it is natural to identify
new constant with Plank's constant $\hbar$ and write $\lambda$ in
the form
\begin{equation} \label{P}
\lambda = \frac{\hbar}{P} ~,
\end{equation}
where quantity $P$ has the dimension of the momentum. Inserting
(\ref{P}) into (\ref{delta-l}) we can conclude that uncertainty
principle
\begin{equation} \label{U}
\Delta x \Delta P > \hbar ~,
\end{equation}
probably has same geometrical meaning as the existence of the
maximal velocity.

To summarize, split quaternions with the norm
\begin{equation} \label{qN}
N = c^2t^2 + \frac{\hbar^2}{P^2} - x^2 - y^2 ~,
\end{equation}
could be used to describe dynamics of a particle with spin in a
2-dimensional space. The two fundamental constants $c$ and $\hbar$
needed for this case have the geometrical origin and correspond to
two kinds of critical signals in (2+2)-space. The Lorentz factor
\begin{equation} \label{qLorentz}
\gamma = \sqrt{1 - \frac{v^2}{c^2} + \frac{\hbar^2}{c^2}~
\frac{F^2}{P^2} }~,
\end{equation}
corresponding to a general boost in the space (\ref{qN}) contains
an extra positive term, where $F$ is some kind of force. Thus, the
dispersion relation in the (2+2)-space (\ref{qN}) has a form
similar to that of double-special relativity models \cite{double}.


\section{Octonions}

Despite the fascination of octonions for over a century since
their discovery in 1844-1845 by  Graves and Cayley, it is fair to
say that they still await universal acceptance. The octonions have
remained in the shadow for a long time. This is not to say that
there have not been various attempts to find appropriate uses for
them in physics (see, for example \cite{SoLo,Qua,Oct,Ku}).
Recently we can say that the octonions have revealed themselves as
the most important number system of all, since they are crucial to
string theory \cite{string}.

As we have seen in previous sections, hyper-numbers are useful to
describe dynamics in 1-dimensional space, and split-quaternions,
in 2-dimensional space. According to the Hurwitz theorem there is
only one further composite division algebra, the octonion algebra
\cite{Sc}. We want to show that split-octonions can describe
dynamics in the usual 3-dimensional space.

For the construction of the octonion algebra with the general
element
\begin{equation} \label{O}
O = a_0 e_0 + a_n e_n ~, ~~~~~n =1,2, ..., 7
\end{equation}
where $e_0$ is the unit element and $a_0, a_n$ are real numbers,
the multiplication law of its eight basis units $e_0, e_n$ usually
is given. For the case of ordinary octonions
\begin{equation} \label{en}
e_0^2 = e_0~,~~ e_n^2 = - e_0~, ~~ e_0^* = e_0 ~, ~~ e_n^* = - e_n
~,
\end{equation}
and norms are positively defined
\begin{eqnarray} \label{NO+}
N = OO^* = O^*O = \nonumber\\
= a_0^2 + a_1^2 + a_2^2 + a_3^2 + a_4^2 + a_5^2 + a_6^2 + a_7^2 ~.
\end{eqnarray}
The multiplication table for the basis elements (\ref{en}) can be
written in the form
\begin{equation} \label{ee}
(e_n e_m) = - \delta_{nm}e_0 + \epsilon_{nmk}\epsilon_k ~,
\end{equation}
where $ \delta_{nm}$ is the Kronecker symbol and $\epsilon_{nmk}$
is the fully anti-symmetric tensor with the postulated value
$\epsilon_{nmk} = + 1$ for the following values of indices
$$
nmk = 123, 145, 176, 246, 257, 347, 365 ~.
$$
This definition of structure constants used by Cayley in his
original paper but is not unique. There are $16$ different
possibilities of definition of $\epsilon_{nmk}$ leading to
equivalent algebras. Sometimes for visualization of the
multiplication law of the octonion basis units the geometrical
picture of the Fano plane (a little gadget with 7 points and 7
lines) also used \cite{Ba}. Unfortunately both of these methods
are almost unenlightening. We want to consider a more obvious
picture using properties of usual vectors products.

For the case of quaternions we had shown that not all basis
elements of the algebra are independent. One easily recover the
full algebra from two fundamental basis elements without
consideration of multiplication tables or graphics. The same can
be done with the octonions. Now beside of unit element $1$ we have
the three fundamental basis elements $i$, $j$ and $l$ of
split-octonions with the properties of ordinary unit vectors
\begin{equation} \label{i2j2l2}
i^2 = j^2 = l^2 = 1~.
\end{equation}
Other four basis elements $(ij)$, $(il)$, $(jl)$ and $((ij)l)$ can
be constructed by vector multiplications of $i$, $j$ and $l$. All
brackets denote vector product and they show only the order of
multiplication. Here we have some ambiguity of the placing of
brackets for the eights element $((ij)l)$. However, we can choose
any convenient form, all the other possibilities are defined by
the ordinary laws of triple vector multiplication. As it was
mentioned in the introduction to denote scalar and vector
multiplications of basis units we does not need to use different
kind of brackets, since the product of the different basis vectors
is always the vector product, and the product of two same basis
vectors is always the scalar product.

Using properties of standard scalar and vector products of $i$,
$j$ and $l$ we can write the properties of other four, not
fundamental, basis elements of the octonion algebra
\begin{eqnarray} \label{ijl}
(i(jl)) = - ((ij)l) = - ((li)j)~, \nonumber\\
(ij) = -(ji)~, ~~~ (il) = -(li) ~, ~~~ (jl) = -(lj) ~, \nonumber \\
(ij)^2 = (il)^2 = (jl)^2 = -1~,  \\
(ij)(ij)^* = (il)(il)^* = (jl)(jl)^* = 1~,\nonumber \\
((ij)l)^2 = 1 ~, ~~~ ((ij)l)((ij)l)^* = -1 ~. \nonumber
\end{eqnarray}
Conjugation as for the quaternion case means reflection of the
basis units, or changing of there signs (except of unit element).
These obvious properties of the scalar and vector products of
basis elements are hidden if we write the abstract octonion
algebra (\ref{ee}).

In the algebra of split-octonions it is possible to introduce four
classes (totally eight) of projection operators
\begin{eqnarray} \label{DO}
D^\pm_i = \frac{1}{2}(1 \pm i)~,~~~~~~~D^\pm_j = \frac{1}{2}(1 \pm
j)~, \nonumber \\
 D^\pm_l = \frac{1}{2}(1 \pm l) ~, ~~
D^\pm_{(ij)l} = \frac{1}{2}(1 \pm (ij)l) ~,
\end{eqnarray}
corresponding to the critical unit signal along $i,j,l$ and
$((ij)l)$ directions. These four classes do not commute with each
other. Only one class, i.e. any pair $D^\pm$ with the same label
from $i, j, l$ and $((ij)l)$, is independent.

In the algebra we have also the four classes (totally twenty four)
of Grassmann numbers labelled by $i, j, l$ and $((ij)l)$
\begin{eqnarray} \label{GGO}
G^\pm_{i1} = \frac{1}{2}(1 \pm i)j ~, ~~~
G^\pm_{i2} = \frac{1}{2}(1 \pm i)l ~, \nonumber\\
G^\pm_{i3} = \frac{1}{2}(1 \pm i)(jl) ~, ~~~~~\nonumber \\
G^\pm_{j1} = \frac{1}{2}(1 \pm j)i ~,~~~
G^\pm_{j2} =\frac{1}{2}(1 \pm j)l ~, \nonumber \\
G^\pm_{j3} = \frac{1}{2}(1 \pm j)(il) ~, ~~~~~\nonumber\\
G^\pm_{l1} = \frac{1}{2}(1 \pm l)i ~, ~~~
G^\pm_{l2} = \frac{1}{2}(1 \pm l)j ~, \\
G^\pm_{l3} = \frac{1}{2}(1 \pm l)(ij) ~, ~~~~~\nonumber\\
G^\pm_{(ij)l1} = \frac{1}{2}(1 \pm (ij)l )i ~, ~~~
G^\pm_{(ij)l 2} = \frac{1}{2}(1 \pm (ij)l)j ~, \nonumber\\
G^\pm_{(ij)l 3} = \frac{1}{2}(1 \pm (ij)l)l ~. ~~~~~\nonumber
\end{eqnarray}
Only the numbers $G_1, G_2$ and $G_3$ from each class with the
same sign mutually commute. So we have eight independent real
Grassmann algebras with the elements $G^\pm_{n1}, G^\pm_{n2}$ and
$ G^\pm_{n3}$, where $n$ runs over $i, j, l$ and $((ij)l)$.

In contrast with quaternions, octonions cannot be represented by
matrices with the usual multiplication lows \cite{Sc}. The reason
is the non-associativity, leading to the different rules for the
left and right multiplications.

Using commuting projection operators and Grassmann numbers any
split octonion
\begin{eqnarray} \label{Os}
O = A + Bi + Cj + D(ij) + \nonumber\\
+ El + F(il) + G(jl) + H(i(jl)) ~,
\end{eqnarray}
where $A, B, C, D, E, F, G$ and $H$ are some real numbers, can be
written in the form
\begin{eqnarray} \label{O1}
O = D^+ [(A + B) + (C + D)G^+_{1} + \nonumber \\
+ (E + F) G^+_{2} + (G + H) G^+_{3}] + \nonumber \\
+ D^- [(A - B) + (C - D)G^-_{1} + \\
+(E - F) G^-_{2}  + (G - H) G^-_{3}] ~. \nonumber
\end{eqnarray}
Because of the symmetry between $i, j$ and $l$ using (\ref{DO})
and (\ref{GGO}) we can find three different representation of
(\ref{O1}).

Two terms in formula (\ref{O1}) with the opposite signs in the
labels of zero divisors correspond to the direct and backward
signals for the one of three real directions $i, j$, or $l$.
However, as for quaternions, the different classes of projection
operators $D_i, D_j$ and $D_l$ do not commute. This is the analog
of non-commutativity of coordinates \cite{Non}.

As in the case with quaternions, here we also have the two
fundamental constants $c$ and $\hbar$ corresponding to the two
types of critical signals $D^\pm$ and $G^\pm$. Thus any octonion
can be written in the form
\begin{eqnarray} \label{Op}
O = ct + xi + yj + zl + \nonumber\\
+\frac{\hbar}{P_z}(ij) + \frac{\hbar}{P_y} (il) +
\frac{\hbar}{P_x} (jl) + \omega (i(jl)) ~,
\end{eqnarray}
where $\omega$ is some quantity with the dimension of length. We
had used expressions similar to (\ref{Nzch}) and (\ref{qN}) to
extract $c$ and $\hbar$ from the components of (\ref{Os}) and to
introduce the quantities $ P_x, P_y$ and $ P_z $ with the
dimensions of momentum.

If we extract the two fundamental constants $c$ and $\hbar$ from
$\omega$, we find a quantity with the dimensions of energy
\begin{equation}\label{E}
E = \frac{c\hbar}{\omega} ~.
\end{equation}
So the norm of the octonion
\begin{eqnarray} \label{ON}
N = OO^* = c^2t^2 + \frac{\hbar^2}{P_x^2}+ \frac{\hbar^2}{P_y^2} +
\frac{\hbar^2}{P_z^2} - \nonumber \\
- x^2 - y^2 - z^2 - \frac{c^2\hbar^2}{E^2} ~,
\end{eqnarray}
becomes similar to a interval in some kind of phase space.

The appearance of Planck's constant in definitions of the
intervals (\ref{ON}) means that the difference of the proposed
model from the standard approach mainly will take place for large
energies and nonlinear fields. For example, the non-associativity
can be connected with Green's functions of strongly non-linear
Quantum Field Theory considered in \cite{Dz}.

Specific to the representation (\ref{O1}) is the existence of
three zero divisors forming full Grassmann algebras with three
elements for any direction. So for octonions we have four type of
commuting critical signals. One corresponds to the rotations in
$(t-x)$ plane giving the constant $c$, two correspond to the
different rotations in $(x-P)$ planes, which give uncertainty
relations similar to (\ref{delta-l}) containing $\hbar$, and the
last one corresponds to the rotations in $(P-E)$ plane.

Note that in spite of the fact that in the split-octonion algebra
there are four real basis vectors $i, j, l$ and $((ij)l)$ we can
consider it to present only three ordinary space dimensions, since
there could be constructed only the three independent rotation
operators with the property $R^+R^- = 1$,
\begin{eqnarray} \label{RO}
R^\pm_{ij} = \frac{1}{\sqrt{2}}[1 \pm (ij)] ~, \nonumber \\
R^\pm_{il} = \frac{1}{\sqrt{2}}[1 \pm (il)]~, \\
R^\pm_{jl} = \frac{1}{\sqrt{2}}[1 \pm (jl)]~, \nonumber
\end{eqnarray}
corresponding to rotations in the $(i-j)$, $(i-l)$ and $(j-l)$
planes respectively. Another fact is that in the formulae
(\ref{DO}) and (\ref{GGO}) there is no symmetry between the zero
divisors along $i, j$, or $l$ and along $((ij)l)$. The reason is
non-associativity of octonions resulting in non-unique answers for
the products of $D^\pm_{(ij)l} $ and $G^\pm_{(ij)l} $. This means
that the commutating relations of these operators depends on the
order of multiplication of the basis elements and we could not
write a unique decomposition (\ref{O1}) of octonion using
$D^\pm_{(ij)l}$ and $G^\pm_{(ij)l}$.

The basis element $((ij)l)$ always appears for the rotations
around $i, j$ and $l$. For example, the rotation by $\pi /2$
around $i$ (i.e. in the $(j-l)$ plane) by the operators (\ref{RO})
gives $((jl)i)$. After a rotation on $\pi $ we receive $-i$ and
only after the full rotation bt $2\pi$ do we come back to $i$,
indeed
\begin{equation} \label{Ri}
R^+_{jl} i R^-_{jl} = -i(jl) = (ij)l ~, ~~~ R^+_{jl} ((ij)l)
R^-_{jl} = -i ~.
\end{equation}
A similar situation occurs for the rotations around $j$ and $l$.

So non-associativity, which results in non-visibility of the
fourth real direction, introduces fundamental uncertainty in
3-space. Again because of non-associativity the extra relations
\begin{equation} \label{delta-e}
\Delta t \Delta E > \hbar ~, ~~~~~\frac{\Delta E}{ \Delta P} >
c\hbar ~,
\end{equation}
in addition to (\ref{delta-x}) and (\ref{delta-l}), following from
the positive definiteness of the norm (\ref{ON}) do not have the
same meaning of the uncertainty relation as in (\ref{U}). However,
(\ref{delta-e}) leads to generalizations of standard uncertainty
relation similar as considered in various approaches \cite{FL}.

Non-associativity of octonions follows from the property of triple
vector products and appears in multiplications of three different
basis elements of octonion necessarily including as the one of the
terms the eighth basis element $((ij)l)$. For example, we have
\begin{eqnarray} \label{nonasso}
j \left( l ~((ij)l)\right) = j \left(-(ij)\right) = i ~, \nonumber\\
(jl)~((ij)l) = (jl)((jl)i) = -i ~.
\end{eqnarray}

Non-associativity of octonions is difficult to understand if we
only study the properties of basis elements. For example, if one
introduce so-called open multiplication laws for the basis
elements \cite{Ku} (i.e. does not place the brackets) and formally
use only the anti-commutation properties , he find
\begin{equation} \label{ijl1}
ijl = lij ~.
\end{equation}
But anti-commuting comes from the properties of vector products
and it is impossible to use it separately. If we put the brackets
back then, from the properties of vector products, we discover the
another result
\begin{equation} \label{(ij)i}
((ij)l) = - (l(ij))~.
\end{equation}
This relation is obvious because expresses the vector product of
two vectors $l$ and $(ij)$.

Non-associativity of octonions is the result of the pseudo-vector
character of the basis elements $(ij), (il)$ and $(jl)$, which
have the properties of an imaginary unit, i.e. there square
(scalar product) is negative. Even in three dimensions the
ordinary vector multiplication is non-associative
\begin{equation} \label{abc}
[[{\bf ab}]{\bf c}] \ne [{\bf a}[{\bf bc}]] ~.
\end{equation}
For example, if ${\bf a}$ and ${\bf b}$ are two unit orthogonal
3-vectors we have
\begin{equation} \label{non-ass}
[[{\bf aa}]{\bf b}] = 0 ~, ~~~ [{\bf a}[{\bf ab}]]=-{\bf b} ~.
\end{equation}
In 3-space the triple vector product
\begin{equation} \label{abc1}
[{\bf a}[{\bf bc}]] = ({\bf ac}){\bf b} - ({\bf ab}){\bf c}
\end{equation}
of three orthogonal vectors (describing by the determinant of
three-on-three matrix) is zero. Considering vector products of
orthogonal vectors in more than three dimensions, it is possible
to construct extra vectors, normal to all three (using matrix with
the more than three columns and rows). Then the triple vector
product of three orthogonal vectors will be not zero and we shall
automatically discover their non-associative properties. In
particular, when we have a product of three different basis
elements of octonion, and when one of them is the eighth basis
unit $((ij)l)$, for some order of multiplication there necessarily
arises the scalar product (square) of two pseudo-vectors from
$(ij), (il)$ and  $(jl)$, which give the opposite sign in the
result.

In general, the familiar 3-dimensional identity for triple vector
product (\ref{abc1}) does not follows from the experimentally
established properties of vectors (\ref{e-vector}). Also, it is
not valid in seven dimensions, the only dimension other than three
where the vector product can be written in principle \cite{Ro,Si}.

Octonions have a weak form of associativity called alternativity.
In the language of basis elements this means that in the
expressions with only two fundamental basis elements the placing
of brackets is arbitrary. For example,
\begin{equation} \label{ii}
((ii)l) = (i(il))~, ~~~ ((li)i) = (l(ii)) ~.
\end{equation}
This property, following from the ordinary properties of vector
products allows us to simplify expressions containing more than
three octonionic unit elements. The rule is: fundamental basis
units of the same type, which after commuting appears to be
neighbors, can be taken out of the brackets to identify there
squares with the unit of the algebra.

At the end we want to note, that in general the relations
(\ref{delta-x}), (\ref{delta-l}) and (\ref{delta-e}) are not need
to be satisfied separately. Positive definiteness of the norm
(\ref{ON}) gives only one relation containing all components of
the octonion. The Lorentz factor corresponding to a general boost
in (4+4)-space has the form
\begin{equation} \label{OLorentz}
\gamma = \sqrt{1 - \frac{v^2}{c^2}  - \hbar^2
~\left(\frac{A^2}{E^2} - \frac{1}{c^2}~ \frac{F^2}{P^2}\right) }~,
\end{equation}
where the quantities $A$ and $F$ have the dimensions of work and
force respectively. The formula (\ref{OLorentz}) contains extra
terms, which change 4-dimensional standard dispersion relation in
the similar way to that of the double-special relativity theory
\cite{double}.


\section{Conclusion}

Using the assumptions that the observed space-time geometry must
follow from the properties of the algebra that we use in physical
measurements and that the result of any measurement is given by
the norm of split-octonionic signals, in this paper it was found
that:

a). In the theory we must have one time (the unit element of
octonion) and three real space directions (corresponding to the
vector-like basis elements $i$,$j$ and $l$).

b). There are two fundamental constants $c$ and $\hbar$ connected
with the projection operators and Grassmann numbers respectively
(critical elements of the algebra).

c). From the positive definiteness of the norms there arise
uncertainty relations and fundamental uncertainty in measuring of
coordinates (because of existence of a fourth space-like direction
with non-associative properties).

We do not yet introduce any physical fields or equations. We have
only shown that the proper algebra of split-octonions could in
principle describe the observed geometry and have introduced many
necessary characteristics of the future particle physics model.


\section*{Acknowledgements}

Author would like to acknowledge the hospitality extended during
his visits at Theoretical Divisions of CERN and SLAC where this
work was done.


\end{document}